\documentclass[preprint,prb,aps,showpacs,floatfix]{revtex4}

\usepackage{graphicx}
\usepackage{dcolumn}
\usepackage{bm}
\usepackage{amssymb}

\def\nut{$\nu_\mathrm{tot}$}
\def\bpa{$B_{||}$}
\def\wpk{$\omega_\mathrm{pk}$}
\def\mwpk{\omega_\mathrm{pk}}
\def\dw{$\Delta\omega$}
\def\vr{$V(\vec{r})$}
\def\vtr{$V^t(\vec{r})$}
\def\vbr{$V^b(\vec{r})$}
\def\mvr{V(\vec{r})}
\def\mvtr{V^t(\vec{r})}
\def\mvbr{V^b(\vec{r})}
\def\sigxx{Re[$\sigma_{xx}$($\omega$)]}
\def\sqt{$\frac{1}{\sqrt{2}}$}
\def\msqt{\frac{1}{\sqrt{2}}}

\def\uket{$|\uparrow\rangle$}
\def\dket{$|\downarrow\rangle$}
\def\dsas{$\Delta_\mathrm{SAS}$}

\begin{document}

\preprint{}

\title{Pinned Bilayer Wigner Crystals with Pseudospin Magnetism}

\author{Yong P. Chen}
\altaffiliation[Current address: ]{Dept. of Physics and Astronomy, Rice University, Houston TX 77005}
\affiliation{Department of Electrical Engineering, Princeton University,
Princeton, NJ 08544}
\affiliation{National High Magnetic Field Laboratory, 1800 E.~Paul Dirac Drive, 
Tallahassee, FL 32310}
\email{yongchen@rice.edu}

\date{\today}

\begin{abstract}
We study a model of \textit{pinned} bilayer Wigner crystals (WC) and focus on the effects of 
interlayer coherence (IC) on pinning.  We consider both a pseudospin ferromagnetic WC (FMWC) with
IC and a pseudospin antiferromagnetic WC (AFMWC) without IC. Our central finding is that a FMWC can be pinned more strongly due to the presence of IC.  One specific mechanism is through the disorder induced interlayer tunneling, which effectively manifests as an extra pinning in a FMWC. We also construct a general ``effective disorder" model and effective pinning Hamiltonian for the case of FMWC and AFMWC respectively. Under this framework, pinning in the presence of IC involves \textit{interlayer} spatial correlation of disorder in addition to intralayer correlation, leading to \textit{enhanced} pinning in the FMWC.   The pinning mode frequency (\wpk) of a FMWC is found to decease with the effective layer separation, whereas for an AFMWC the opposite behavior is expected.  An abrupt drop of \wpk\ is predicted at a transition from a FMWC to AFMWC.  Possible effects of in-plane magnetic fields and finite temperatures are addressed.  Finally we discuss some other possible ramifications of the FMWC as an electronic supersolid-like phase.
\end{abstract}
\maketitle

\section{Introduction}
Two dimensional systems (2DS) of electrons (or holes) 
subjected to a strong perpendicular magnetic field ($B$) have been among the most studied strongly-correlated systems in the past two decades,  with such 
many-body phenomena as fractional quantum Hall effects and quantum Wigner crystals (WC) (both reviewed in Ref.~\onlinecite{dass:97}).  
Additional degrees of freedom introduced by bringing two parallel 2DS in close separation to form a bilayer system (BLS) can lead to new phenomena (see reviews in 
Refs.~\onlinecite{dass:97,simo:05}) with no counterpart in the single layer case. The best known example is the bilayer excitonic condensate state\cite{fert:89,wenz:92} (BECS) at total Landau filling \nut=1, which displays quantum Hall effect\cite{murp:94} and counterflow superfluidity\cite{kell:04,tutu:04}.  Carriers in such a state reside \textit{simultaneously} in both layers and possess interlayer (phase) coherence (IC). The IC can even exist in the limit of vanishing interlayer tunneling (characterized by the symmetric-antisymmetric energy gap \dsas\ $\sim$0) and solely due to the interlayer Coulomb interaction\cite{simo:05}.  Alternatively, the IC can be described using a pseudospin \cite{yang:94,moon:95} language, where pseudospins represent layer indices.  The BECS is a pseudospin ferromagnet\cite{yang:94,moon:95} and the associated Goldstone mode has indeed been observed\cite{spie:01}.  

At sufficiently small \nut, the ground state of the BLS is expected to be a bilayer Wigner crystal\footnote{WC-type phases have also been considered at \nut=1 as possible competing phases
\cite{brey:90,xche:92,cote:92,deml:01,veil:02,yang:01}with BECS at intermediate layer separation.} (BWC)\cite{oji:87}.  It is natural to ask whether IC can also exist in the crystal state.  Such a possibility has been theoretically considered\cite{zhen:95,nara:95}, for finite as well as vanishing interlayer tunneling. It was found\cite{zhen:95,nara:95} that when $d/a$ (the effective layer separation, where $d$ is the interlayer spacing and $a$ is the mean intralayer spacing between carriers)
is small, the BWC can be an one-component Wigner crystal with IC. This corresponds to a WC which is also a pseudospin easy-plane ferromagnet\cite{nara:95}.   For larger $d/a$, on the other hand, the BWC is expected to be a two-component WC (TCWC)\cite{nara:95}. The two components 
(corresponding to the two layers) are ``staggered" from each other in order to minimize interlayer Coulomb interaction.   If interlayer tunneling is small, such a TCWC has negligible IC and is an easy-axis antiferromagnet\footnote{With finite tunneling, it was shown\cite{nara:95} that the TCWC can have a mixed ferromagnetic-antiferromagnetic order with a (small) net pseudospin magnetization.} in pseudospin space.  
A rich array of crystal structures\cite{zhen:95,nara:95,esfa:95,ggol:96b} was shown to be possible with a TCWC other than the standard hexagonal lattice\cite{bons:77}.  Dynamical properties of a BWC have
been calculated\cite{ggol:96b,falk:94,klir:04}.

So far theories\cite{zhen:95,nara:95,esfa:95,ggol:96b,falk:94,klir:04} on BWC have focused on the clean case.  However, in real samples a BWC is always pinned by disorder and is therefore an insulating phase as observed in experiments\cite{mano:96,tutu:03}.  Disorder can also introduce 
a pinning gap in the magnetophonon excitation of a WC\cite{fuku:78,norm:92}. Such a ``pinning mode"\cite{fert:99,fogl:00,chit:02} has been taken as a well-defined characteristic signature for a 
pinned WC measured in the single layer case\cite{ccli:97,henn:98,pdye:02,chen:04}. 

In this article we study \textit{pinned} BWC and in particular, we focus on the effect of interlayer
coherence (or pseudospin magnetism) on the pinning mode and experimentally detectable signatures
 that can \textit{qualitatively} distinguish a pseudospin ferromagnetic (FM) WC from a pseudospin antiferromagnetic (AFM) WC in real bilayer systems.

After a brief review of the pinning mode in a single layer (SL) WC in Sec.~\ref{sec:singlepin}, 
we develop a simple model of pinned BWC in Sec.~\ref{sec:bipin} to calculate the 
pinning mode properties both with and without IC.  First we demonstrate that local tunneling induced by
disorder (such as barrier fluctuations) manifests as an effective pinning in the presence of IC and 
can lead to \textit{enhanced} pinning in the FMWC. Then we present a more general model, where the concept of \textit{effective disorder}, which depends on the electronic state, is emphasized. 
Under this framework, pinning in the presence of IC involves \textit{interlayer} as well as intralayer spatial correlation of disorder, whereas only the latter is relevant for the pinning for a SL WC or a BWC without IC.    The effect of $d/a$ on the pinning mode frequency (\wpk) is discussed in Sec.~\ref{sec:da}. Qualitatively opposite behaviors are found for a FMWC and an AFMWC. We also predict an abrupt \wpk\ drop associated with a FM-to-AFM transition.  In Sec.~\ref{sec:bieff} we discuss possible effects of in-plane magnetic fields (\bpa) on a pinned FMWC. A proposal of performing ``disorder tomography" using \bpa\  is presented. We also briefly discuss finite-temperature ($T$) effects.   In Sec.~\ref{sec:supersol} we discuss some other interesting properties (and their connection with pinning) of FMWC as a phase resembling a supersolid. We summarize the paper in Sec.~\ref{sec:bipsum}.

\section{Pinning of a Single Layer WC\label{sec:singlepin}}

In the presence of disorder\footnote{Provided it is sufficiently ``weak",
which perturbs (deforms) but does not destroy the WC.}, a WC cannot have true 
long range positional order\cite{chit:02}.  Its long wavelength and low energy excitation is the 	``pinning  
mode"\cite{fuku:78,norm:92,fert:99,fogl:00,chit:02}, which represents the collective oscillation of WC
domains in the disorder potential.   Such a pinning mode is manifested as a resonance in the frequency-dependent real diagonal conductivity (Re[$\sigma_{xx}$($\omega$)]), measurable from the power absorption spectrum of the WC subjected to an AC electric field\cite{ccli:97,pdye:02}.  Major results from the current understanding of the pinning mode resonance are summarized below, where we consider a 
WC with density $n$ subject to a (weak) disorder potential $V(\vec{r})$ (where $\vec{r}$ denotes the position vector in the 2D plane) and a strong perpendicular $B$: 
\begin{description}
\item[(i)]
The frequency of the pinning mode resonance (\wpk) is only determined by the \textit{static} deformation
(from the ideal lattice in the clean case) of the WC through its Larkin domain size \cite{fert:99,fogl:00,chit:02}.  
An explicit formula for \wpk (in the high $B$ limit) as given in Ref.~\onlinecite{chit:02} is\footnote{Different theories\cite{fert:99,fogl:00,chit:02} so far differ on the exponent of $\xi$ appearing in Eq.~(\ref{eq:wpk}) but this, as will be seen, is unimportant for our purposes.}
\begin{equation}
\omega_\mathrm{pk}=C\frac{W}{\xi^6}\frac{1}{\mu}\frac{1}{B}  \label{eq:wpk}
\end{equation}
In this formula $C$ is a constant involving only the carrier charge ($e$), $\mu$ is the shear modulus of the WC, $W$ and $\xi$ are the strength and correlation length of the (effective) disorder
(see \textbf{(v)} below) potential ($V(\vec{r})$). They are defined from the two-point spatial correlator
\begin{equation}
\langle V(\vec{r})V(\vec{r}')\rangle=WD_\xi(|\vec{r}-\vec{r}'|)  \label{eq:disco}
\end{equation}
where $D_\xi(r)$ is the correlation function with characteristic decay length $\xi$.  
 
 For an ideally 2D (infinitely thin) WC in high $B$, $\mu$ is expected to be close to its classical 
value\cite{bons:77} 
\begin{equation}
\mu=\alpha\frac{n^{3/2}e^2}{\epsilon}
\end{equation}
where $\epsilon$ is the effective dielectric constant of the medium and $\alpha$
a constant set by the crystal structure ($\sim$0.02 for the hexagonal lattice). 
Thus the expected $n$-dependence of \wpk\ is 
\begin{equation}
\mwpk\propto n^{-\gamma}
\end{equation}
with $\gamma$=3/2.    Experimentally measured\cite{ccli:00} $\gamma$ varies from 1/2 to 3/2.
Its precise value is not qualitatively important for this work. 
\item[(ii)]
The determination of the line width ($\Delta \omega$) of the pinning mode resonance is less 
straightforward.  It is now believed \cite{fert:99,fogl:00} to be a truly dynamical quantity and determined by the magnetophonon localization length.  In general, (at a fixed $B$), $\Delta \omega$ increases with increasing disorder, but decreases with increasing Coulomb interaction strength.
\item[(iii)]
The integrated intensity ($S$) of the pinning mode resonance directly reflects the participating density
of the WC. It is shown\cite{fuku:78} that $S=(ne/4B)\mwpk$. 
\item[(iv)] It has been suggested\cite{fert:99} that the physical disorder responsible for the pinning comes mainly from the roughness associated with the interface that vertically confines the WC.  
Such disorder gives rise to a calculated\cite{fert:99} \wpk\  comparable to that observed experimentally\cite{ccli:97,pdye:02}.
\item[(v)] Although the physical disorder is assumed to not to depend on the electronic state, the \textit{effective} disorder (\vr) which determines the Wigner crystal pinning, is electronic state dependent.   More specifically, \vr\ is the physical disorder appropriately \textit{convoluted} with
the electron form factor (wave function)\cite{fert:99,fogl:00,chit:02} .  As a consequence,  the disorder correlation length $\xi$ appearing in (\ref{eq:wpk}) above is that of the physical disorder ($\xi_0$)
only when $\xi_0 > l_B$ (valid at sufficiently high $B$), where the magnetic length $l_B=\sqrt{\hbar/eB}$ is the size of one electron wave function.  Otherwise (if $\xi_0 < l_B$), $\xi$ should be set as $l_B$. 
\end{description}

\section{Pinned Bilayer Wigner Crystals with Pseudospin Magnetism\label{sec:bipin}}
Now consider a BWC of equal densities ($n$) of electrons in each layer, 
with interlayer separation $d$ and in a strong perpendicular $B$. 
We assume the disorder in the ``top" layer (\vtr) and that in the ``bottom" layer (\vbr) to 
be similar\footnote{Rigorously, we are assuming that \vtr\ and \vbr\ are two \textit{realizations} of the same \textit{random field} \vr\label{ft:dis}.}: 
\begin{equation}
\mvtr\sim\mvbr\sim\mvr   \label{eq:bidis}
\end{equation} 
where \vr\ obeys the disorder characteristics defined in Eq.~(\ref{eq:disco}) and already incorporates 
the appropriate \textit{intralayer} electron form factor. Therefore, in the absence of the other layer, each would form a SL pinned WC with the same pinning mode as described in Sec.~\ref{sec:singlepin}. 
In the following we will use the superscripts
 ``$n0$'' to denote quantities associated with the pinning mode of 
such a SL WC (of density $n$), and ``$nn$" for those associated with the BWC (of total densities $2n$, with $n$ in each layer).  We use $N$ to denote 
the number of electrons in each layer ($N$=$nA$ with $A$ being the sample area) and pseudospin ``$\uparrow$" and  ``$\downarrow$" for ``top"  and ``bottom"  layer indices respectively (we also assume
both layers to be infinitely thin, located at $z$=$+$$d/2$ and $z$=$-$$d/2$ respectively, where
$(x,y,z)$=$(\vec{r},z)$ are 3D Cartesian coordinates for a 2D (intralayer) vector $\vec{r}$). We can 
ignore the real spin degree of freedom for electrons in high $B$ (the lowest Landau level).

Our model can be presented clearly in the first quantized language. We start with the \textit{total}
Hamiltonian for the pinned BWC 
\begin{equation}
\hat{H}=\sum_{i=1}^{N}(\hat{H}_s(\vec{r}_i,\uparrow)+\hat{H}_s(\vec{r}_i,\downarrow))+
\hat{U}_\mathrm{int}
+\hat{V}_\mathrm{dis} \label{eq:biHam}
\end{equation}
In the above 
$\hat{H}_s$ is the single-particle part of the Hamiltonian, which also includes a neutralizing positive charge background (to keep the total Coulomb energy finite) but does \textit{not} include disorder effects.
 
The Coulomb interaction among all electrons is
\begin{eqnarray}
\hat{U}_\mathrm{int}&=&\sum_{i<j}^{N}\frac{e^2}{|(\vec{r}_i,\uparrow)-(\vec{r}_j,\uparrow)|}+\sum_{i<j}^{N}\frac{e^2}{|(\vec{r}_i,\downarrow)-(\vec{r}_j,\downarrow)|} \nonumber\\
&&+\sum_{i,j}^{N}\frac{e^2}{|(\vec{r}_i,\uparrow)-(\vec{r}_j,\downarrow)|} \label{eq:biCol}
\end{eqnarray}
in which the first two terms represent intralayer interaction, the third term represents interlayer interaction, with $|(\vec{r}_i,\uparrow)-(\vec{r}_j,\uparrow)|=|(\vec{r}_i,\downarrow)-(\vec{r}_j,\downarrow)|=|\vec{r}_i-\vec{r}_j|$ and $|(\vec{r}_i,\uparrow)-(\vec{r}_j,\downarrow)|=\sqrt{|\vec{r}_i-\vec{r}_j|^2+d^2}$.

The disorder part, $\hat{V}_\mathrm{dis}$, has two parts $\hat{V}_\mathrm{dis}=\hat{V}_\mathrm{pin}+\hat{T}_\mathrm{dis}$.   One is the pinning within each layer
\begin{equation}
\hat{V}_\mathrm{pin}=\sum_{i=1}^{N}(V^t(\vec{r}_i,\uparrow)+V^b(\vec{r}_i,\downarrow)) \label{eq:bipin}
\end{equation}

Note we have explicitly written out the configuration space coordinates above ((\ref{eq:biHam})-
(\ref{eq:bipin})) to reflect its layer (pseudospin) dependent actions.  For example,
\begin{equation}
V(\vec{r},\uparrow)|\psi(\vec{r})\otimes\downarrow\rangle=V(\vec{r},\downarrow)|\psi(\vec{r})\otimes
\uparrow\rangle=0 \label{eq:disact}
\end{equation} 
for a single-particle state $\psi$, where the notation $|\psi(\vec{r})\otimes
\rho\rangle$ ($\rho$ being a pseudospin state) is a shorthand for the state $|\psi\rangle\otimes|\rho\rangle$.

The other part in $\hat{V}_\mathrm{dis}$ reflects effect
of disorder induced (local) tunneling and is given by
\begin{equation} 
\hat{T}_\mathrm{dis}=T(\vec{r})\hat{F}     \label{eq:dtun}
\end{equation}
where $\hat{F}$ is simply the pseudospin flip operator 
\begin{equation} 
\hat{F}|\psi(\vec{r})\otimes\downarrow\rangle=|\psi(\vec{r})\otimes\uparrow\rangle, \hat{F}|\psi(\vec{r})\otimes\uparrow\rangle=|\psi(\vec{r})\otimes\downarrow\rangle
\end{equation}
and the amplitude $T(\vec{r})$ is generally related to \vtr\ and \vbr.

We first notice that if there were no interlayer coupling (for example $d\gg$$a$$\sim$$1/\sqrt{n}$),
both the interlayer interaction term in (\ref{eq:biCol}) and disorder induced tunneling (\ref{eq:dtun}) 
can be neglected and $\hat{H}$ decouples into
two identical (only shifted in $z$) SL Hamiltonians.  In this case the system reduces
to two independent layers and its pinning mode resonance (\sigxx\ spectrum) is simply 
the superposition of those of two identical SL WC, i.e, \sigxx$^{nn}$=$2$\sigxx$^{n0}$
with $\mwpk^{nn}$= $\mwpk^{n0}$, $\Delta\omega$$^{nn}$=$\Delta\omega$$^{n0}$ and $S^{nn}$=2$S^{n0}$ in this independent layer limit.

In this article we are mainly interested in \textit{interacting} bilayers and we focus on the effect
of IC on the pinning mode of a BWC, in particular on \wpk, which is the quantity that can be most accurately measured in experiments\cite{ccli:97,pdye:02}.  To this end, we will consider and compare two idealized cases of a BWC with \textit{no} IC (referred to as an ``AFMWC") and a BWC with IC (``FMWC"), to be specified by the many-body ansatz (\ref{eq:afmwc}) and (\ref{eq:fmwc}) 
 in the following respectively.   Our approach is to construct an \textit{effective} Hamiltonian that 
maps the problem into a single layer one, with an \textit{effective} disorder that captures the pinning physics in each case (AFMWC \textit{vs} FMWC),  highlighting the difference made by IC,  and calculate quantities such as \wpk. 

We also make the following additional assumptions,  which greatly simplify the analysis but still keep the essential physics. 
\begin{enumerate}
\item  Assume small or vanishing interlayer tunneling ($\Delta_\mathrm{SAS}$) in absence of disorder.  We also assume the neutralizing positive charges are far away from the BWC. Together with the high $B$ condition (which allows us to neglect the cyclotron kinetic energy of the electrons),  the $\hat{H}_s$ part in $\hat{H}$ is nearly constant and can be neglected all together.  Physically, this means that pinning is only determined by the 
electron-electron interaction ($\hat{U}_\mathrm{int}$) and electron-disorder interaction ($\hat{V}_\mathrm{dis}$): the static deformation (which determines pinning \wpk, as we have pointed out in Sec.~\ref{sec:singlepin}) is given by 
the configuration that minimizes the energy expectation of $\hat{U}_\mathrm{int}+\hat{V}_\mathrm{dis}$.
Although $\hat{V}_\mathrm{dis}$ contains both disorder pinning and disorder tunneling parts, we will see later on in the article, that (in either case of an AFMWC or a FMWC) $\hat{V}_\mathrm{dis}$ can be replaced by some \textit{effective} pinning potential in a simpler \textit{effective} Hamiltonian (which does not contain a disorder tunneling term) that gives the same \wpk\ as the original problem.
\item  Assume $d\ll a$.  This in particular allows us to effectively set $d$$\sim$0 in the Coulomb
interaction term ($\hat{U}_\mathrm{int}$) in (\ref{eq:biCol}) and treat the inter and intralayer interactions
on an equal footing.  In this limit we can also assume the underlying lattice structure (in absence of disorder) to be the same (hexagonal) for the AFMWC and FMWC\cite{nara:95}.
\item  Assume the following simple form for the disorder induced tunneling amplitude: 
\begin{equation}
T(\vec{r})=\tilde{g}V(\vec{r}) \label{eq:gvr}
\end{equation}
 where $\tilde{g}\ge 0$ is a small (we only consider the effect of disorder being weak perturbation) parameter.  This is plausible because we expect the main source of relevant disorder in realistic, epitaxially-grown samples to come from the defects or fluctuations in the thin barrier separating the two layers.  This kind of barrier defects or fluctuations not only constitute disorder in each layer, they can also facilitate local tunneling\cite{tutu:04,shay:co} between the layers, with a tunneling amplitude being locally proportional to the strength of such disorder (in the weak disorder limit) as expressed in 
 (\ref{eq:gvr}).  The positive sign of $\tilde{g}$ comes from the fact that such a tunneling-facilitating defect draws an electron closer into the barrier and farther away 
from the corresponding positive charged background/dopants, therefore constituting a positive disorder. We also expect $\tilde{g}$ to decrease with increasing effective layer separation $d/a$ and go to zero at large $d/a$ (the decrease of $\tilde{g}$ with decreasing $a$ reflects the Coulomb-blocking effect on the tunneling).
\end{enumerate}
Later on we will briefly discuss the implication when the above
assumptions are relaxed, which nonetheless will not change our qualitative conclusions. 

\begin{figure}
\includegraphics[width=9cm]{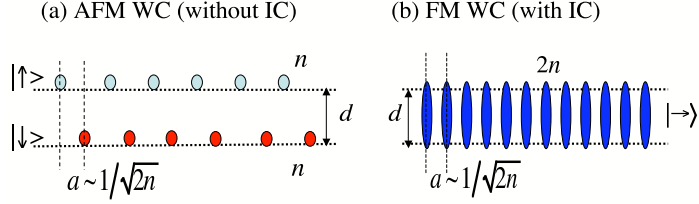}
\caption{\label{fig:BWC} Schematic (1D cross section, not to scale, $d$ assumed to be $\ll$$a$) for an AFMWC (a) and a FMWC (b). Both have total densities $2n$ and the same underlying lattice structure (when not deformed by disorder). In (a) half the lattice electrons belong to the top layer (labeled as $\uparrow$) and the other half to the bottom layer
(labeled as $\downarrow$). Electrons are only pinned by disorder from the individual layer. In (b) all electrons \textit{simultaneously} belong to both layers (being in pseudospin state $|\rightarrow\rangle$=
\sqt\uket+\sqt\dket), and are effectively pinned by the \textit{joint}  disorder (see the text for details) from both layers. }
\end{figure}

\underline{\textit{Case 1.} AFMWC (no IC).} 

A schematic picture (1D cross section) is shown in Fig.~\ref{fig:BWC}(a).  This corresponds to a ``bipartite"\cite{nara:95} lattice \{$\vec{R}_i$\}$^{2N}_{i=1}$ (deformed slightly from the ideal lattice \{$\vec{R}^0_i$\}$^{2N}_{i=1}$). We have relabeled the
indices such that $i=1,\ldots,N$ correspond to the ``$\uparrow$" electrons and $i=N+1,\ldots,2N$ correspond to ``$\downarrow$" electrons.  The many-body state of the AFMWC can be well approximated  by the following ansatz\cite{nara:95,maki:83} (after appropriate antisymmetrization)
\begin{equation}
\Psi_\mathrm{AFMWC}=\prod^{2N}_{i=1}|\psi_{\vec{R}_i}(\vec{r}_i)\otimes\rho_i\rangle \label{eq:afmwc}
\end{equation}
in which 
\begin{equation}
\rho_i=\uparrow \mathrm{\ for\ } i=1,\ldots,N \mbox{\ and\ } \downarrow \mathrm{\ for\ } i=N+1,\ldots,2N  \label{eq:afmrho}
\end{equation}  
and the single-particle Gaussian (up to a phase)\\
 $\psi_{\vec{R}}(\vec{r})$=$\frac{1}{\sqrt{2\pi}l_B}\exp[-\frac{|\vec{r}-\vec{R}|^2}{4l_B^2}]
\exp[-\mathrm{i}\frac{\hat{\mathrm{z}}\cdot(\vec{r}\times\vec{R})}{2l_B^2}$] (where
 $l_B$=$\sqrt{\hbar/eB}$ is the magnetic length and $\hat{\mathrm{z}}$ the unit $z$-vector.)  

With our index relabeling (and assumption $d\ll a$) we can rewrite 
\begin{equation}
\hat{U}_\mathrm{int}=\sum_{i<j}^{2N}\frac{e^2}{|\vec{r}_i-\vec{r}_j|}
\end{equation}
Following Eqs.~(\ref{eq:bidis},\ref{eq:bipin},\ref{eq:disact},\ref{eq:afmwc},\ref{eq:afmrho}) we easily see 
\begin{equation}
\hat{V}_\mathrm{pin}|\Psi_\mathrm{AFMWC}\rangle\sim\sum_{i=1}^{2N}V(\vec{r}_i)
|\Psi_\mathrm{AFMWC}\rangle
\end{equation}
We also notice that, for the pseudospin flip operator $\hat{F}$,
$\langle\psi(\vec{r})\otimes\rho |\hat{F}|\psi(\vec{r})\otimes\rho\rangle$=0 for a single particle state $\psi(\vec{r})$ and $\rho$=either $\uparrow$ or $\downarrow$. Therefore in the case of an AFMWC, the disorder induced tunneling $\hat{T}_\mathrm{dis}$ does \textit{not} affect the pinning (static deformation) of the crystal\footnote{In a first order approximation (with small $\tilde{g}$), we can neglect the effect of $\hat{T}$ modifying the spin structure of the crystal.}.  

The \textit{effective} Hamiltonian for the \textit{pinned} AFMWC is 
\begin{equation}
\hat{H}^{pin}_\mathrm{AFMWC}=
\sum_{i<j}^{2N}\frac{e^2}{|\vec{r}_i-\vec{r}_j|}+\sum_{i=1}^{2N}V(\vec{r}_i) \label{eq:afmpin}
\end{equation}
where the pseudospins have dropped out.  Thus as far as pinning is concerned, the system 
maps to a \textit{SL} of $2N$ electrons crystallizing in the effective disorder potential \vr.
The static deformation of such a crystal can be 
obtained in principle by minimizing the energy with respect to \{$\vec{R}_i$\}$^{2N}_{i=1}$, using the many-body ansatz $\Psi_\mathrm{AFMWC}$ (\ref{eq:afmwc}) with this effective Hamiltonian 
(\ref{eq:afmpin}). 
For its pinning mode we simply have (from Sec.~\ref{sec:singlepin}) 
\begin{equation}
 \mwpk^{nn}= \mwpk^{n0}/2^\gamma, \mathrm{\ } \Delta\omega^{nn}<\Delta\omega^{n0}, 
 \mathrm{\ and\ }S^{nn}=S^{n0}/2^{\gamma-1}  \label{eq:afmpinmd}
\end{equation}

\underline{\textit{Case 2.} FMWC (with IC).}
A schematic (1D cross section) is shown in Fig.~\ref{fig:BWC}(b).  In contrast to \textit{Case. 1}, whose
lattice is bipartite with the AFM order, here the lattice is one-component with all electrons in the
pseudospin state $|\rightarrow\rangle$ (=\sqt$|\uparrow\rangle$+\sqt$|\downarrow\rangle$).
Such  FM order breaks the U(1) symmetry\footnote{The full SU(2) symmetry of pseudospins is already 
broken \textit{explicitly} by the bilayer capacitive charging energy.} of pseudospins (either \textit{explicitly} by finite \dsas\
 or \textit{spontaneously} (only due to interlayer Coulomb interaction) for \dsas$\sim$0). The many-body
ansatz for such a FMWC is
\begin{equation}
\Psi_\mathrm{FMWC}=\prod^{2N}_{i=1}|\psi_{\vec{R}_i}(\vec{r}_i)\otimes\rightarrow\rangle \label{eq:fmwc}
\end{equation} (where $\psi_{\vec{R}}(\vec{r})$ is the same kind of Gaussian wavepacket used earlier).

Compared to \textit{Case 1}, now $\hat{T}_\mathrm{dis}$ has a very different effect:
it is easy to see that 
\begin{equation}
\hat{T}_\mathrm{dis}(\vec{r})|\psi(\vec{r})\otimes\rightarrow\rangle=\tilde{g}\mvr|\psi(\vec{r})\otimes\rightarrow\rangle
\end{equation}
for a single particle state $\psi(\vec{r})$.   This means that, in contrast to the case of AFMWC (without IC), 
where $\hat{T}_\mathrm{dis}$ does not affect pinning as seen earlier,  the disorder induced tunneling $\hat{T}_\mathrm{dis}$ in the presence of IC effectively acts as a pinning term (this in fact holds even for 
a general tunneling disorder (\ref{eq:dtun})). In our case, this pinning is in addition to the original 
``intra-layer"  pinning from \vr,  thus leads to enhanced pinning of a FMWC (\ref{eq:fmwc}).   

Now we construct an alternative, \textit{effective} disorder model in which the system is mapped into 2$N$ electrons crystallizing in a single  ``$\rightarrow$" layer,  and pinning effects such as that due to $\hat{T}_\mathrm{dis}$ above are absorbed in an effective (single layer) pinning disorder,
given by the following ``joint" disorder\footnote{One may explicit write $V^J$ as $V^J(\vec{r},\rightarrow)$ to emphasize that it is an \textit{effective disorder} acting on ``$|\rightarrow\rangle$" electrons. $V^J$ may also be formally thought as resulting from a kind of ``convolution" in pseudospin space.  However, a rigorous definition of such a convolution requires appropriately defining an invariant measure on a pseudospin algebra, and is beyond the scope of this paper.} ansatz (the reason for the choice will be soon apparent):
\begin{equation}
V^J(\vec{r})=\msqt(\mvtr+\mvbr)   \label{eq:jdis}
\end{equation}
with the \textit{effective} pinning Hamiltonian being
\begin{equation}
\hat{H}^{pin}_\mathrm{FMWC}=
\sum_{i<j}^{2N}\frac{e^2}{|\vec{r}_i-\vec{r}_j|}+\sum_{i=1}^{2N}V^J(\vec{r}_i)
\end{equation}

The spatial correlator for such a ``joint" disorder $V^J$ now contains (in terms of the original bilayers) both intralayer (Eq.~(\ref{eq:disco})) and \textit{interlayer disorder-correlation}:
\begin{eqnarray}
\lefteqn{\langle V^J(\vec{r})V^J(\vec{r}')\rangle}\nonumber\\
&=&\frac{1}{2}[\langle V^t(\vec{r})V^t(\vec{r}')\rangle+\langle V^b(\vec{r})V^b(\vec{r}')\rangle\nonumber\\
&&+\langle V^t(\vec{r})V^b(\vec{r}')\rangle+\langle V^b(\vec{r})V^t(\vec{r}')\rangle]\nonumber\\
&=&W(1+g)D_\xi(|\vec{r}-\vec{r}'|)
\end{eqnarray}
in which we have introduced a phenomenological ``coupling" parameter $g$ between the disorder from
the two layers: 
\begin{equation}
 \langle V^t(\vec{r})V^b(\vec{r}')\rangle=\langle V^b(\vec{r})V^t(\vec{r}')\rangle=gWD_\xi(|\vec{r}-\vec{r}'|)  \label{eq:interdis}
\end{equation}
Again we expect $g$ to depend on the effective layer separation ($d/a$): $g$ decreases for increasing $d/a$ and drops to 0 at sufficiently large $d/a$.  Now
we see that $V^J$ has disorder strength $W^J$=$(1+g)W$ and the same correlation length ($\xi$) as 
\vr.  Thus we obtain for the bilayer pinning mode properties 
(expressed in terms of corresponding SL ``$n0$" quantities):
\begin{equation}
\mwpk^{nn}=\frac{1+g}{2^\gamma}\mwpk^{n0}, \mathrm{\ \ \  } S^{nn}=\frac{1+g}{2^{\gamma -1}}S^{n0}
\label{eq:fmwcpm}
\end{equation}
In contrast, the interlayer disorder-correlation (\ref{eq:interdis}) has no relevance for pinning of the AFMWC (\ref{eq:afmpin}-\ref{eq:afmpinmd}) or the SL (``$n0$") WC 
(Sec.~\ref{sec:singlepin}). Therefore, although the doubled $n$ (and strengthened Coulomb interaction) in the BWC (``$nn$") from the SL (``$n0$") case will decrease \wpk\  and \dw;  in a FMWC,  the presence of IC effectively \textit{enhances} the pinning disorder and tends to increase both \wpk\  and \dw\  from the respective SL (``$n0$") values.  Due to the two competing effects, \wpk\  (\ref{eq:fmwcpm}) and \dw\  for the FMWC can be \textit{either} higher or lower than the $\mwpk^{n0}$ and \dw$^{n0}$.  In contrast, \wpk\  and \dw\  for the AFMWC are \textit{always} lower than the SL  values. Detailed calculations\cite{chen:up} (following Ref.~\onlinecite{chit:02}) show that for the FMWC, if  $\mwpk^{nn}$=$\mwpk^{n0}$, \dw$^{nn}$$<$\dw$^{n0}$.

The ``effective disorder" (in the effective Hamiltonian) model we give above does not directly specify the source of the inter-layer correlated disorder (such as barrier fluctuations) with the enhanced pinning mechanism.  However it correctly captures (now in the disorder correlator, which mathematically determines \wpk\ as shown in Eqs.~(\ref{eq:wpk},\ref{eq:disco})) the effects such interlayer disorder may have on pinning; furthermore, through the choice of the ``joint" disorder (\ref{eq:jdis}), it carries a simple physical picture that, in the state of FMWC, since electrons have lost their original layer identity and move in both layers simultaneously and coherently, they are pinned by disorder from both layers.  Such a general framework
turns out to be convenient to analyze the BWC pinning properties in Sec.~\ref{sec:da} and \ref{sec:bieff}
below.  

\section{Effects of $d/a$ and FMWC-AFMWC transitions\label{sec:da}}
\begin{figure}
\includegraphics[width=9cm]{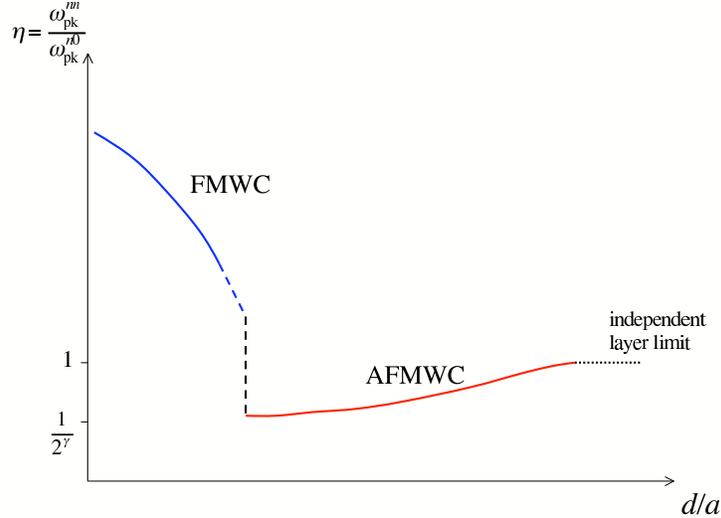}
\vspace{1cm}
\caption{\label{fig:da} Schematic ($d/a$)-dependence of $\eta$ (BWC $\mwpk^{nn}$ normalized by SL   
$\mwpk^{n0}$), showing an \textit{abrupt} FMWC to AFMWC transition (characterized by 
a sudden drop in $\mwpk^{nn}$) and a continuous AFMWC to independent-layer cross over.  The asymptotic values of $\eta$ for $d/a$$\rightarrow$0 in AFMWC and $d/a$$\rightarrow$$\infty$ (independent layer limit) are $1/2^\gamma$ and 1 respectively (see Sec.~\ref{sec:bipin}).}
\end{figure}

As seen from the above, for the FMWC, \wpk\ will \textit{decrease} when the effective layer separation $\delta$(=$d/a$) increases, due to the decrease of $g$ (Eq.~(\ref{eq:fmwcpm})).  It has been shown\cite{zhen:95,nara:95} that at some small critical $\delta_c$, a transition from
a FMWC (favored at $\delta$$<$$\delta_c$) to an AFMWC (favored at $\delta$$>$$\delta_c$) occurs.  
Since pinning in the AFMWC (without IC) does not involve $g$, such a transition would result in a sudden reduction of pinning and would give rise to an abrupt drop of \wpk\  (see Fig.~\ref{fig:da}, in which we plot
the schematic dependence of \wpk\ (normalized by the SL $\mwpk^{n0}$) on $d/a$).  

If $\delta$ is further increased (in an AFMWC) such that $d$ becomes comparable to or even 
larger than $a$, the interlayer Coulomb interaction will be reduced. This reduces the total Coulomb 
interaction (\ref{eq:biCol}) and effectively reduces the shear modulus ($\mu$) of the BWC. From  (\ref{eq:wpk}), this will give rise to an \textit{increase} of \wpk.  In the limit of
$d$$\gg$$a$, the system reduces to two independent SL WC and $\mwpk^{nn}/\mwpk^{n0}\rightarrow 1$.  

Thus we have shown (Fig.~\ref{fig:da}) that \wpk\ can have \textit{opposite} behavior in the FMWC 
(\wpk\ decreasing with increasing $d/a$) from that in the AFMWC (\wpk\ increasing with 
increasing $d/a$), and \wpk\ drops \textit{abruptly} at a FMWC-AFMWC transition.  Such behavior 
can qualitatively differentiate a FMWC from an AFMWC and signal the transition between the two.  

It was found earlier\cite{nara:95,zhen:95} that if tunneling (\dsas) is finite, the BWC at $\delta$$>$$\delta_c$, although two-component, can have mixed AFM-FM order,  corresponding\cite{nara:95} to $|\rho_i\rangle$=$|\nearrow\rangle$ ($i$=1,...,$N$) and $|\rho_i\rangle$=$|\searrow\rangle$ ($i$=$N$+1,...,2$N$) in (\ref{eq:afmwc}), where (pseudospin direction)``$\nearrow$"(``$\searrow$") 
is ``$\uparrow$"(``$\downarrow$") tilted toward ``$\rightarrow$" by angle $\theta$ ($\theta$=0 for an ideal AFMWC considered so far).
$\theta$=$\pi/2$ for the FMWC ($\delta$$<$$\delta_c$) and \textit{drops abruptly} to a finite value 
(0$<$$\theta$$<$$\pi/2$) at the transition (at $\delta_c$) \cite{nara:95}. Therefore we 
expect the abrupt drop of \wpk\  associated with the transition to survive even with a moderate 
\dsas\  , although the \textit{amplitude} of the 
drop will be smaller than the \dsas$\sim$0 case. 

At finite $\delta$$>$$\delta_c$, Ref.~\onlinecite{nara:95} also found several possible lattice structures (without disorder) and a continuous evolution among them  as a function of $\delta$.  The evolution is gradual
and is not expected to change the qualitative picture shown in Fig.~\ref{fig:da}, in particular the presence of the abrupt drop of \wpk\ at the FM-AFM transition. 

Since the enhancement of pinning in the FMWC is associated with the presence of IC, we expect the abrupt \wpk\ drop to be a generic feature whenever IC (or equivalently, ferromagnetism) is 
destroyed, even if it is driven by some other mechanisms (such as changing 
$\nu_\mathrm{tot}$\cite{nara:95}, or possibly with sufficient layer imbalance\cite{tutu:03,chen:up}).
  
\section{Effects of In-Plane Magnetic Fields and Finite Temperatures\label{sec:bieff}}
\begin{figure}
\includegraphics[width=8.5cm]{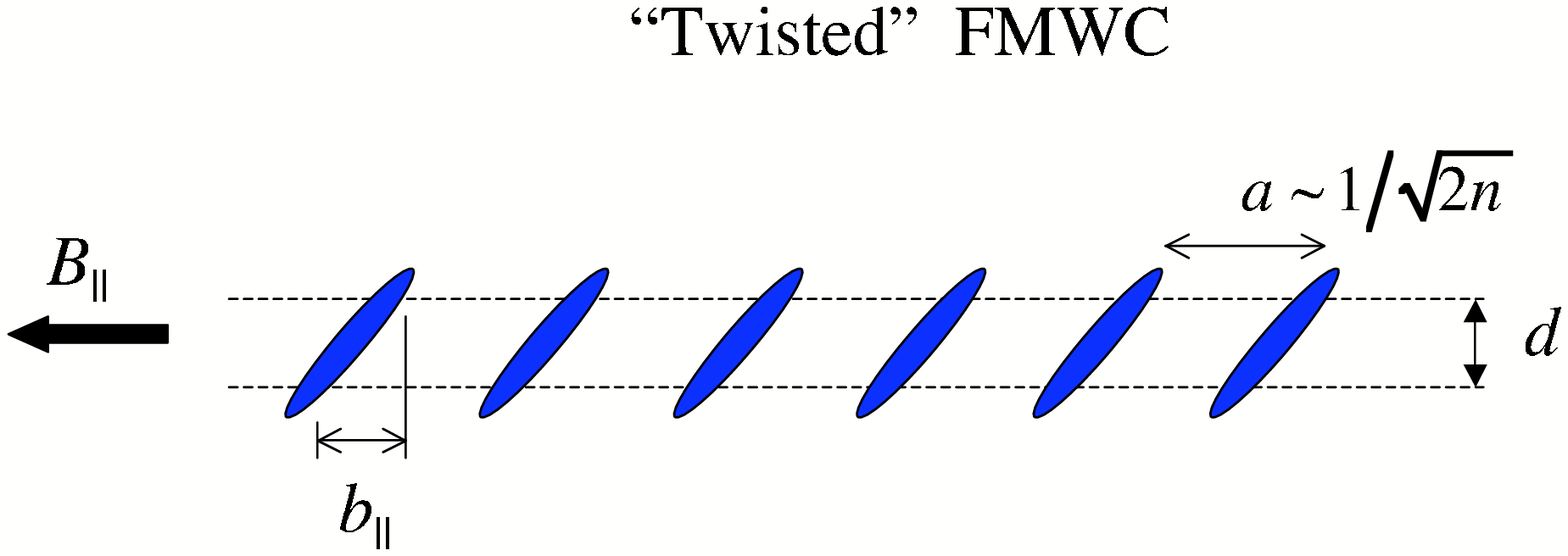}
\caption{\label{fig:bpa} A FMWC is ``twisted" under a (small) in-plane magnetic field (\bpa), which can also ``twist" the IC-induced interlayer disorder coupling (now involving  $\langle V^t(\vec{r})V^b(\vec{r}'+\vec{b}_{||})\rangle$, where $V^t$($V^b$) is the top(bottom) layer disorder.  
The ``twist" $\vec{b}_{||}$=$d$$(B_{||}/B)$$\hat{\mathrm{x}}$, where $\hat{\mathrm{x}}$ is the direction of \bpa. }
\end{figure}

\underline{\textit{In-plane magnetic fields} ($B_{||}$)}. It is well known that $B_{||}$ can profoundly affect bilayer physics\cite{yang:94}, 
particularly in relation to interlayer phase coherence and pseudospin magnetism.   In the case of a bilayer FMWC (with finite $d$), Zheng and Fertig\cite{zhen:95} studied the effects of $B_{||}$ and found that applying a \textit{small} \bpa\ can ``twist" the IC, such that the charge distribution in 
one layer is shifted relative to the other layer, as shown in Fig.~\ref{fig:bpa}.  The relative shift is along the \bpa\ direction ($\hat{\mathrm{x}}$), and is given\cite{zhen:95} by 
$\vec{b}_{||}=l^2_{B}(d/l^2_{B_{||}})\hat{\mathrm{x}}=d(B_{||}/B)\hat{\mathrm{x}}$. 
In such a case, the interlayer disorder coupling induced by IC can also become ``twisted" \cite{fert:co},
now involving interlayer disorder-correlation 
\begin{equation}
\langle V^t(\vec{r})V^b(\vec{r}'+\vec{b}_{||})\rangle=gWD_\xi(|\vec{r}-\vec{r}'-\vec{b}_{||}|)
\end{equation}
and this will duly affect the pinning.  Therefore, measuring \wpk\ while varying both the direction and magnitude of \bpa\  allows one to possibly probe a 2D ``tomography" of the disorder!

At larger \bpa\ , an incommensurability-driven transition to an ``untwisted" state  
is expected to occur, when the energy cost of interlayer Coulomb interaction exceeds the energy gain from interlayer hopping \cite{zhen:95}.  We expect such a transition to cause also an abrupt change of
\wpk\ in the pinning mode.

\underline{\textit{Finite temperatures} ($T$)}.  So far we have considered only $T$=0.  
Finite $T$ is expected 
in particular to smear the abrupt drop in \wpk\ associated with the FMWC-AFMWC transition as described in Sec.~\ref{sec:da}.  Above some characteristic $T$ ($T_*$), such a drop would become unobservable.  The typical energy difference between a FMWC and AFMWC has been shown \cite{zhen:95,nara:95} to be on the order of $\Delta E\sim$$10^{-3}$$-$$10^{-2}e^2/\epsilon l_B$.  From this, we can make a (very rough) estimate of $T_*$ to be on the order of $\Delta E/k_B$$\sim$0.5K (using a typical experimental $l_B$$\sim$100\AA\  and $\epsilon$=13 for GaAs)\footnote{Also assuming the BWC has not melted at such $T$.}.    

\section{FMWC as a Supersolid-like Phase\label{sec:supersol}}
Finally we remark on two interesting aspects of the 
FMWC (\ref{eq:fmwc}), particularly in the case of vanishing \dsas, 
and speculate on effects in relation to pinning.    We may rewrite the many-body state (\ref{eq:fmwc})
in second-quantized form as 
\begin{equation}
\Psi_\mathrm{FMWC}=\frac{1}{\sqrt{2}}\prod_{i}(c^\dag_{\vec{R}_i,\uparrow}
+e^{\mathrm{i}\phi}c^\dag_{\vec{R}_i,\downarrow})|0\rangle  \label{eq:fmwc2}
\end{equation}
with $\phi=0 \mathrm{\  for\ all\ } i$, where $c^\dag_{\vec{r},\downarrow}$ ($c^\dag_{\vec{r},\uparrow}$) is the second-quantized operator that creates an electron localized at $\vec{r} $\   in down (up) layer.
The ansatz (\ref{eq:fmwc2}) is formally analogous to that 
of the bilayer excitonic condensate state (BECS) \cite{simo:05,fert:89} ($\frac{1}{\sqrt{2}}\prod_{i}(c^\dag_{\vec{k}_i,\uparrow}
+c^\dag_{\vec{k}_i,\downarrow})|0\rangle$) with single-particle states labeled with 
momenta ($\vec{k}_i$) replaced by those with lattice point positions ($\vec{R}_i$). 
The FMWC (\ref{eq:fmwc2}) possesses both long range positional order (broken translational symmetry) and phase ($\phi$) coherence (broken U(1) symmetry), thus resembles a \textit{supersolid}\cite{legg:70,yang:co} phase.  If phase stiffness (associated with $\phi$) exists, superflow would occur and would be exhibited in the counterflow channel, similar to the case observed in the BECS\cite{kell:04,tutu:04}.  Although such a superflow in the FMWC is likely 
to be suppressed by the pinning (at least in a linear response theory),  it would be an interesting
experiment to examine the counterflow with finite current above the \textit{depinning} threshold,
or under a sufficiently strong parallel magnetic field (\bpa) which can reduce the effective pinning 
associated with IC or disorder induced interlayer tunneling\footnote{Disorder in tunneling is 
expected to be particularly detrimental to counterflow superfluidity\cite{xwen:co}. A parallel magnetic field can suppress interlayer tunneling\cite{zhen:95}.} described in (\ref{eq:dtun}).   
 
The Goldstone mode associated with the (spontaneously) broken U(1) symmetry represents an oscillatory wave in $\phi$. 
Since $\phi$ is conjugate to the density difference between the two layers ($\Delta n$),  such a U(1) mode inevitably involves interlayer charge transfer (oscillation in $\Delta n$) and will 
be coupled to the longitudinal and transverse phonons (which are hybridized in $B$) of the WC\cite{klir:04}.  We expect that such coupling to the U(1) mode (which disperses linearly in $k$) can
not only \textit{renormalize} the pinning gap (with enhanced pinning, as we have seen), but also
the dispersion of the pinning mode (as the lowest lying hybridized mode)\footnote{Analogous couplings have been studied for the helium supersolid, see for example Ref.~\onlinecite{ycch:81}.}. Such a dispersion change may be detectable in the $k$-resolved microwave spectroscopy experiments\cite{henn:98,chen:04} and also be used to identify the FMWC phase.    
 
\section{Conclusion\label{sec:bipsum}}

BWC can display a rich array of (pseudospin) quantum magnetism from FM to 
AFM order \cite{nara:95}. They are in many ways analogous to  $^3$He solid\cite{adam:04}, which  
has many remarkable physical properties related to its quantum magnetism and the FMWC may even be 
considered as an electronic supersolid-like phase.  In this article we have 
focused on the effects of pseudospin magnetism on the pinning by disorder, which always
exists in a real BWC.   Electrons in a FMWC have interlayer coherence (IC) and lose their individual layer identities, similar to the situation in the $\nu_\mathrm{tot}=1$  quantum Hall state.   We have shown that such IC can take advantage of the interlayer correlation of disorder (such as through disorder 
in the barrier and the interlayer tunneling induced by such disorder) and \textit{enhance} the effective pinning in the FMWC.  The IC-enhanced pinning is a novel mechanism that has no counterpart in a single layer WC and is absent in an AFMWC without IC.   
For the pinning mode resonance, this has important consequences
which may be used as experimental signatures of the different magnetic phases and phase transitions in BWC.  For example, we predict  \wpk\ to decrease with $d/a$ in a FMWC but to increase with $d/a$ in a AFMWC, with an \textit{abrupt} drop of \wpk\ at a FMWC-AFMWC transition.   We have also considered effects of \bpa\ and finite temperatures.   Many predictions of our model are found to be consistent with a recent experimental work by Z.~Wang \textit{et al.} \cite{zwan:to}.
\vspace{1cm} 
 
\begin{acknowledgments}
The author thanks Herb Fertig, Kun Yang, Leonid Levitov, Xiao-Gang Wen and Mansour Shayegan 
for many valuable comments and discussions, as well as Lloyd Engel and Dan Tsui 
for encouragement, and Zhihai Wang for making his experimental data available prior to publication. 
The work is supported by NSF, DOE and AFOSR.
\end{acknowledgments}


\begin{thebibliography}{46}
\expandafter\ifx\csname natexlab\endcsname\relax\def\natexlab#1{#1}\fi
\expandafter\ifx\csname bibnamefont\endcsname\relax
  \def\bibnamefont#1{#1}\fi
\expandafter\ifx\csname bibfnamefont\endcsname\relax
  \def\bibfnamefont#1{#1}\fi
\expandafter\ifx\csname citenamefont\endcsname\relax
  \def\citenamefont#1{#1}\fi
\expandafter\ifx\csname url\endcsname\relax
  \def\url#1{\texttt{#1}}\fi
\expandafter\ifx\csname urlprefix\endcsname\relax\def\urlprefix{URL }\fi
\providecommand{\bibinfo}[2]{#2}
\providecommand{\eprint}[2][]{\url{#2}}

\bibitem[{\citenamefont{{Das Sarma} and Pinczuk}(1997)}]{dass:97}
\bibinfo{editor}{\bibfnamefont{S.}~\bibnamefont{{Das Sarma}}} \bibnamefont{and}
  \bibinfo{editor}{\bibfnamefont{A.}~\bibnamefont{Pinczuk}}, eds.,
  \emph{\bibinfo{title}{Perspectives in Quantum Hall Effects}}
  (\bibinfo{publisher}{Wiley and Sons}, \bibinfo{year}{1997}).

\bibitem[{\citenamefont{Simon}(2005)}]{simo:05}
\bibinfo{author}{\bibfnamefont{S.~H.} \bibnamefont{Simon}},
  \bibinfo{journal}{Solid State Commun.}
  \textbf{\bibinfo{volume}{\textbf{134}}}, \bibinfo{pages}{81}
  (\bibinfo{year}{2005}).

\bibitem[{\citenamefont{Fertig}(1989)}]{fert:89}
\bibinfo{author}{\bibfnamefont{H.~A.} \bibnamefont{Fertig}},
  \bibinfo{journal}{Phys.~Rev.~B.} \textbf{\bibinfo{volume}{\textbf{40}}},
  \bibinfo{pages}{1087} (\bibinfo{year}{1989}).

\bibitem[{\citenamefont{Wen and Zee}(1992)}]{wenz:92}
\bibinfo{author}{\bibfnamefont{X.-G.} \bibnamefont{Wen}} \bibnamefont{and}
  \bibinfo{author}{\bibfnamefont{A.}~\bibnamefont{Zee}},
  \bibinfo{journal}{Phys.~Rev.~Lett.} \textbf{\bibinfo{volume}{\textbf{69}}},
  \bibinfo{pages}{1811} (\bibinfo{year}{1992}).

\bibitem[{\citenamefont{Murphy et~al.}(1994)\citenamefont{Murphy, Eisenstein,
  Boebinger, Pfeiffer, and West}}]{murp:94}
\bibinfo{author}{\bibfnamefont{S.~Q.} \bibnamefont{Murphy}},
  \bibinfo{author}{\bibfnamefont{J.~P.} \bibnamefont{Eisenstein}},
  \bibinfo{author}{\bibfnamefont{G.~S.} \bibnamefont{Boebinger}},
  \bibinfo{author}{\bibfnamefont{L.~N.} \bibnamefont{Pfeiffer}},
  \bibnamefont{and} \bibinfo{author}{\bibfnamefont{K.~W.} \bibnamefont{West}},
  \bibinfo{journal}{Phys.~Rev.~Lett.} \textbf{\bibinfo{volume}{\textbf{72}}},
  \bibinfo{pages}{728} (\bibinfo{year}{1994}).

\bibitem[{\citenamefont{Kellogg et~al.}(2004)\citenamefont{Kellog, Eisenstein,
  Pfeiffer, and West}}]{kell:04}
\bibinfo{author}{\bibfnamefont{M.}~\bibnamefont{Kellogg}},
  \bibinfo{author}{\bibfnamefont{J.~P.} \bibnamefont{Eisenstein}},
  \bibinfo{author}{\bibfnamefont{L.~N.} \bibnamefont{Pfeiffer}},
  \bibnamefont{and} \bibinfo{author}{\bibfnamefont{K.~W.} \bibnamefont{West}},
  \bibinfo{journal}{Phys.~Rev.~Lett.} \textbf{\bibinfo{volume}{\textbf{93}}},
  \bibinfo{pages}{036801} (\bibinfo{year}{2004}).

\bibitem[{\citenamefont{Tutuc et~al.}(2004)\citenamefont{Tutuc, Shayegan, and
  Huse}}]{tutu:04}
\bibinfo{author}{\bibfnamefont{E.}~\bibnamefont{Tutuc}},
  \bibinfo{author}{\bibfnamefont{M.}~\bibnamefont{Shayegan}}, \bibnamefont{and}
  \bibinfo{author}{\bibfnamefont{D.~A.} \bibnamefont{Huse}},
  \bibinfo{journal}{Phys.~Rev.~Lett.} \textbf{\bibinfo{volume}{\textbf{93}}},
  \bibinfo{pages}{036802} (\bibinfo{year}{2004}).

\bibitem[{\citenamefont{Yang et~al.}(1994)\citenamefont{Yang, Moon, Zheng,
  MacDonald, Girvin, Yoshioka, and Zhang}}]{yang:94}
\bibinfo{author}{\bibfnamefont{K.}~\bibnamefont{Yang}},
  \bibinfo{author}{\bibfnamefont{K.}~\bibnamefont{Moon}},
  \bibinfo{author}{\bibfnamefont{L.}~\bibnamefont{Zheng}},
  \bibinfo{author}{\bibfnamefont{A.~H.} \bibnamefont{MacDonald}},
  \bibinfo{author}{\bibfnamefont{S.~M.} \bibnamefont{Girvin}},
  \bibinfo{author}{\bibfnamefont{D.}~\bibnamefont{Yoshioka}}, \bibnamefont{and}
  \bibinfo{author}{\bibfnamefont{S.}~\bibnamefont{Zhang}},
  \bibinfo{journal}{Phys.~Rev.~Lett.} \textbf{\bibinfo{volume}{\textbf{72}}},
  \bibinfo{pages}{732} (\bibinfo{year}{1994}).

\bibitem[{\citenamefont{Moon et~al.}(1995)\citenamefont{Moon, Mori, Yang,
  Girvin, MacDonald, Zheng, Yoshioka, and Zhang}}]{moon:95}
\bibinfo{author}{\bibfnamefont{K.}~\bibnamefont{Moon}},
  \bibinfo{author}{\bibfnamefont{H.}~\bibnamefont{Mori}},
  \bibinfo{author}{\bibfnamefont{K.}~\bibnamefont{Yang}},
  \bibinfo{author}{\bibfnamefont{S.~M.} \bibnamefont{Girvin}},
  \bibinfo{author}{\bibfnamefont{A.~H.} \bibnamefont{MacDonald}},
  \bibinfo{author}{\bibfnamefont{L.}~\bibnamefont{Zheng}},
  \bibinfo{author}{\bibfnamefont{D.}~\bibnamefont{Yoshioka}}, \bibnamefont{and}
  \bibinfo{author}{\bibfnamefont{S.-C.} \bibnamefont{Zhang}},
  \bibinfo{journal}{Phys.~Rev.~B.} \textbf{\bibinfo{volume}{\textbf{51}}},
  \bibinfo{pages}{5138} (\bibinfo{year}{1995}).

\bibitem[{\citenamefont{Spielman et~al.}(2001)\citenamefont{Spielman,
  Eisenstein, Pfeiffer, and West}}]{spie:01}
\bibinfo{author}{\bibfnamefont{I.~B.} \bibnamefont{Spielman}},
  \bibinfo{author}{\bibfnamefont{J.~P.} \bibnamefont{Eisenstein}},
  \bibinfo{author}{\bibfnamefont{L.~N.} \bibnamefont{Pfeiffer}},
  \bibnamefont{and} \bibinfo{author}{\bibfnamefont{K.~W.} \bibnamefont{West}},
  \bibinfo{journal}{Phys.~Rev.~Lett.} \textbf{\bibinfo{volume}{\textbf{87}}},
  \bibinfo{pages}{036803} (\bibinfo{year}{2001}).

\bibitem[{\citenamefont{Oji et~al.}(1987)\citenamefont{Oji, MacDonald, and
  Girvin}}]{oji:87}
\bibinfo{author}{\bibfnamefont{H.~C.~A.} \bibnamefont{Oji}},
  \bibinfo{author}{\bibfnamefont{A.~H.} \bibnamefont{MacDonald}},
  \bibnamefont{and} \bibinfo{author}{\bibfnamefont{S.~M.}
  \bibnamefont{Girvin}}, \bibinfo{journal}{Phys.~Rev.~Lett.}
  \textbf{\bibinfo{volume}{\textbf{58}}}, \bibinfo{pages}{824}
  (\bibinfo{year}{1987}).

\bibitem[{\citenamefont{Zheng and Fertig}(1995)}]{zhen:95}
\bibinfo{author}{\bibfnamefont{L.}~\bibnamefont{Zheng}} \bibnamefont{and}
  \bibinfo{author}{\bibfnamefont{H.~A.} \bibnamefont{Fertig}},
  \bibinfo{journal}{Phys.~Rev.~B.} \textbf{\bibinfo{volume}{\textbf{52}}},
  \bibinfo{pages}{12282} (\bibinfo{year}{1995}).

\bibitem[{\citenamefont{Narasimhan and Ho}(1995)}]{nara:95}
\bibinfo{author}{\bibfnamefont{S.}~\bibnamefont{Narasimhan}} \bibnamefont{and}
  \bibinfo{author}{\bibfnamefont{T.}~\bibnamefont{Ho}},
  \bibinfo{journal}{Phys.~Rev.~B.} \textbf{\bibinfo{volume}{\textbf{64}}},
  \bibinfo{pages}{12291} (\bibinfo{year}{1995}).

\bibitem[{\citenamefont{Esfarjani and Kawazoe}(1995)}]{esfa:95}
\bibinfo{author}{\bibfnamefont{K.}~\bibnamefont{Esfarjani}} \bibnamefont{and}
  \bibinfo{author}{\bibfnamefont{Y.}~\bibnamefont{Kawazoe}},
  \bibinfo{journal}{J.~Phys.~C} \textbf{\bibinfo{volume}{\textbf{7}}},
  \bibinfo{pages}{7217} (\bibinfo{year}{1995}).

\bibitem[{\citenamefont{Goldoni and Peeters}(1996)}]{ggol:96b}
\bibinfo{author}{\bibfnamefont{G.}~\bibnamefont{Goldoni}} \bibnamefont{and}
  \bibinfo{author}{\bibfnamefont{F.~M.} \bibnamefont{Peeters}},
  \bibinfo{journal}{Phys.~Rev.~B.} \textbf{\bibinfo{volume}{\textbf{53}}},
  \bibinfo{pages}{4591} (\bibinfo{year}{1996}).

\bibitem[{\citenamefont{Bonsall and Maradudin}(1977)}]{bons:77}
\bibinfo{author}{\bibfnamefont{L.}~\bibnamefont{Bonsall}} \bibnamefont{and}
  \bibinfo{author}{\bibfnamefont{A.~A.} \bibnamefont{Maradudin}},
  \bibinfo{journal}{Phys.~Rev.~B.} \textbf{\bibinfo{volume}{\textbf{15}}},
  \bibinfo{pages}{1959} (\bibinfo{year}{1977}).

\bibitem[{\citenamefont{Falko}(1994)}]{falk:94}
\bibinfo{author}{\bibfnamefont{V.~I.} \bibnamefont{Falko}},
  \bibinfo{journal}{Phys.~Rev.~B.} \textbf{\bibinfo{volume}{\textbf{49}}},
  \bibinfo{pages}{7774} (\bibinfo{year}{1994}).

\bibitem[{\citenamefont{Klironomos and Dorsey}(2004)}]{klir:04}
\bibinfo{author}{\bibfnamefont{F.~D.} \bibnamefont{Klironomos}}
  \bibnamefont{and} \bibinfo{author}{\bibfnamefont{A.~T.}
  \bibnamefont{Dorsey}}, 
   \bibinfo{journal}{Phys.~Rev.~B.} \textbf{\bibinfo{volume}{\textbf{71}}},
  \bibinfo{pages}{155331} (\bibinfo{year}{2005}).

\bibitem[{\citenamefont{Manoharan et~al.}(1996)\citenamefont{Manoharan, Suen,
  Santos, and Shayegan}}]{mano:96}
\bibinfo{author}{\bibfnamefont{H.~C.} \bibnamefont{Manoharan}},
  \bibinfo{author}{\bibfnamefont{Y.~W.} \bibnamefont{Suen}},
  \bibinfo{author}{\bibfnamefont{M.~B.} \bibnamefont{Santos}},
  \bibnamefont{and} \bibinfo{author}{\bibfnamefont{M.}~\bibnamefont{Shayegan}},
  \bibinfo{journal}{Phys.~Rev.~Lett.} \textbf{\bibinfo{volume}{\textbf{77}}},
  \bibinfo{pages}{1813} (\bibinfo{year}{1996}).

\bibitem[{\citenamefont{Tutuc et~al.}(2003)\citenamefont{Tutuc, Melinte, {De
  Poortere}, Pillarisetty, and Shayegan}}]{tutu:03}
\bibinfo{author}{\bibfnamefont{E.}~\bibnamefont{Tutuc}},
  \bibinfo{author}{\bibfnamefont{S.}~\bibnamefont{Melinte}},
  \bibinfo{author}{\bibfnamefont{E.~P.} \bibnamefont{{De Poortere}}},
  \bibinfo{author}{\bibfnamefont{R.}~\bibnamefont{Pillarisetty}},
  \bibnamefont{and} \bibinfo{author}{\bibfnamefont{M.}~\bibnamefont{Shayegan}},
  \bibinfo{journal}{Phys.~Rev.~Lett.} \textbf{\bibinfo{volume}{\textbf{91}}},
  \bibinfo{pages}{076802} (\bibinfo{year}{2003}).

\bibitem[{\citenamefont{Fukuyama and Lee}(1978)}]{fuku:78}
\bibinfo{author}{\bibfnamefont{H.}~\bibnamefont{Fukuyama}} \bibnamefont{and}
  \bibinfo{author}{\bibfnamefont{P.~A.} \bibnamefont{Lee}},
  \bibinfo{journal}{Phys.~Rev.~B.} \textbf{\bibinfo{volume}{\textbf{18}}},
  \bibinfo{pages}{6245} (\bibinfo{year}{1978}).

\bibitem[{\citenamefont{Normand et~al.}(1992)\citenamefont{Normand, Littlewood,
  and Millis}}]{norm:92}
\bibinfo{author}{\bibfnamefont{B.~G.~A.} \bibnamefont{Normand}},
  \bibinfo{author}{\bibfnamefont{P.~B.} \bibnamefont{Littlewood}},
  \bibnamefont{and} \bibinfo{author}{\bibfnamefont{A.~J.}
  \bibnamefont{Millis}}, \bibinfo{journal}{Phys.~Rev.~B.}
  \textbf{\bibinfo{volume}{\textbf{46}}}, \bibinfo{pages}{3920}
  (\bibinfo{year}{1992}).

\bibitem[{\citenamefont{Fertig}(1999)}]{fert:99}
\bibinfo{author}{\bibfnamefont{H.~A.} \bibnamefont{Fertig}},
  \bibinfo{journal}{Phys.~Rev.~B.} \textbf{\bibinfo{volume}{\textbf{59}}},
  \bibinfo{pages}{2120} (\bibinfo{year}{1999}).

\bibitem[{\citenamefont{Fogler and Huse}(2000)}]{fogl:00}
\bibinfo{author}{\bibfnamefont{M.~M.} \bibnamefont{Fogler}} \bibnamefont{and}
  \bibinfo{author}{\bibfnamefont{D.~A.} \bibnamefont{Huse}},
  \bibinfo{journal}{Phys.~Rev.~B.} \textbf{\bibinfo{volume}{\textbf{62}}},
  \bibinfo{pages}{7553} (\bibinfo{year}{2000}).

\bibitem[{\citenamefont{Chitra et~al.}(2002)\citenamefont{Chitra, Giamarchi,
  and {Le Doussal}}}]{chit:02}
\bibinfo{author}{\bibfnamefont{R.}~\bibnamefont{Chitra}},
  \bibinfo{author}{\bibfnamefont{T.}~\bibnamefont{Giamarchi}},
  \bibnamefont{and} \bibinfo{author}{\bibfnamefont{P.}~\bibnamefont{{Le
  Doussal}}}, \bibinfo{journal}{Phys.~Rev.~B.}
  \textbf{\bibinfo{volume}{\textbf{65}}}, \bibinfo{pages}{035312}
  (\bibinfo{year}{2002}).

\bibitem[{\citenamefont{Li et~al.}(1997)}]{ccli:97}
\bibinfo{author}{\bibfnamefont{C.~C.} \bibnamefont{Li}} \bibnamefont{et~al.},
  \bibinfo{journal}{Phys.~Rev.~Lett.} \textbf{\bibinfo{volume}{\textbf{79}}},
  \bibinfo{pages}{1353} (\bibinfo{year}{1997}).

\bibitem[{\citenamefont{Hennigan et~al.}(1998)\citenamefont{Hennigan, Beya,
  Mellor, Ga\'{a}l, Williams, and Henini}}]{henn:98}
\bibinfo{author}{\bibfnamefont{P.~F.} \bibnamefont{Hennigan}},
  \bibinfo{author}{\bibfnamefont{A.}~\bibnamefont{Beya}},
  \bibinfo{author}{\bibfnamefont{C.~J.} \bibnamefont{Mellor}},
  \bibinfo{author}{\bibfnamefont{R.}~\bibnamefont{Ga\'{a}l}},
  \bibinfo{author}{\bibfnamefont{F.~I.~B.} \bibnamefont{Williams}},
  \bibnamefont{and} \bibinfo{author}{\bibfnamefont{M.}~\bibnamefont{Henini}},
  \bibinfo{journal}{Physica B} \textbf{\bibinfo{volume}{\textbf{249}}},
  \bibinfo{pages}{53} (\bibinfo{year}{1998}).

\bibitem[{\citenamefont{Ye et~al.}(2002)\citenamefont{Ye, Engel, Tsui, Lewis,
  Pfeiffer, and West}}]{pdye:02}
\bibinfo{author}{\bibfnamefont{P.~D.} \bibnamefont{Ye}},
  \bibinfo{author}{\bibfnamefont{L.~W.} \bibnamefont{Engel}},
  \bibinfo{author}{\bibfnamefont{D.~C.} \bibnamefont{Tsui}},
  \bibinfo{author}{\bibfnamefont{R.~M.} \bibnamefont{Lewis}},
  \bibinfo{author}{\bibfnamefont{L.~N.} \bibnamefont{Pfeiffer}},
  \bibnamefont{and} \bibinfo{author}{\bibfnamefont{K.~W.} \bibnamefont{West}},
  \bibinfo{journal}{Phys.~Rev.~Lett.} \textbf{\bibinfo{volume}{\textbf{89}}},
  \bibinfo{pages}{176802} (\bibinfo{year}{2002}).

\bibitem[{\citenamefont{Chen et~al.}(2004)\citenamefont{Chen, Lewis, Engel,
  Tsui, Ye, Wang, Pfeiffer, and West}}]{chen:04}
\bibinfo{author}{\bibfnamefont{Y.~P.} \bibnamefont{Chen}},
  \bibinfo{author}{\bibfnamefont{R.~M.} \bibnamefont{Lewis}},
  \bibinfo{author}{\bibfnamefont{L.~W.} \bibnamefont{Engel}},
  \bibinfo{author}{\bibfnamefont{D.~C.} \bibnamefont{Tsui}},
  \bibinfo{author}{\bibfnamefont{P.~D.} \bibnamefont{Ye}},
  \bibinfo{author}{\bibfnamefont{Z.~H.} \bibnamefont{Wang}},
  \bibinfo{author}{\bibfnamefont{L.~N.} \bibnamefont{Pfeiffer}},
  \bibnamefont{and} \bibinfo{author}{\bibfnamefont{K.~W.} \bibnamefont{West}},
  \bibinfo{journal}{Phys.~Rev.~Lett.} \textbf{\bibinfo{volume}{\textbf{93}}},
  \bibinfo{pages}{206805} (\bibinfo{year}{2004}).

\bibitem[{\citenamefont{Li et~al.}(2000)}]{ccli:00}
\bibinfo{author}{\bibfnamefont{C.~C.} \bibnamefont{Li}} \bibnamefont{et~al.},
  \bibinfo{journal}{Phys.~Rev.~B.} \textbf{\bibinfo{volume}{\textbf{61}}},
  \bibinfo{pages}{10905} (\bibinfo{year}{2000}).

\bibitem[{\citenamefont{Shayegan}()}]{shay:co}
\bibinfo{author}{\bibfnamefont{M.}~\bibnamefont{Shayegan}},
  \bibinfo{howpublished}{private communication}.

\bibitem[{\citenamefont{Maki and Zotos}(1983)}]{maki:83}
\bibinfo{author}{\bibfnamefont{K.}~\bibnamefont{Maki}} \bibnamefont{and}
  \bibinfo{author}{\bibfnamefont{X.}~\bibnamefont{Zotos}},
  \bibinfo{journal}{Phys.~Rev.~B.} \textbf{\bibinfo{volume}{\textbf{28}}},
  \bibinfo{pages}{4349} (\bibinfo{year}{1983}).

\bibitem[{\citenamefont{Chen}()}]{chen:up}
\bibinfo{author}{\bibfnamefont{Y.~P.} \bibnamefont{Chen}},
  \bibinfo{howpublished}{unpublished}.

\bibitem[{\citenamefont{Fertig}()}]{fert:co}
\bibinfo{author}{\bibfnamefont{H.~A.} \bibnamefont{Fertig}},
  \bibinfo{howpublished}{private communication}.

\bibitem[{\citenamefont{Leggett}(1970)}]{legg:70}
\bibinfo{author}{\bibfnamefont{A.~J.} \bibnamefont{Leggett}},
  \bibinfo{journal}{Phys.~Rev.~Lett.} \textbf{\bibinfo{volume}{\textbf{25}}},
  \bibinfo{pages}{1543} (\bibinfo{year}{1970}).

\bibitem[{\citenamefont{Yang}()}]{yang:co}
\bibinfo{author}{\bibfnamefont{K.}~\bibnamefont{Yang}},
  \bibinfo{howpublished}{private communication}.

\bibitem[{\citenamefont{Adams}(2004)}]{adam:04}
\bibinfo{author}{\bibfnamefont{E.~D.} \bibnamefont{Adams}},
  \bibinfo{journal}{J.~Low.~Temp.~Phys.}
  \textbf{\bibinfo{volume}{\textbf{135}}}, \bibinfo{pages}{695}
  (\bibinfo{year}{2004}).

\bibitem[{\citenamefont{Wang et~al.}()}]{zwan:to}
\bibinfo{author}{\bibfnamefont{Z.}~\bibnamefont{Wang}} \bibnamefont{et~al.},
  \bibinfo{howpublished}{unpublished}.

\bibitem[{\citenamefont{Brey}(1990)}]{brey:90}
\bibinfo{author}{\bibfnamefont{L.}~\bibnamefont{Brey}},
  \bibinfo{journal}{Phys.~Rev.~Lett.} \textbf{\bibinfo{volume}{\textbf{65}}},
  \bibinfo{pages}{903} (\bibinfo{year}{1990}).

\bibitem[{\citenamefont{Chen and J.Quinn}(1992)}]{xche:92}
\bibinfo{author}{\bibfnamefont{X.~M.} \bibnamefont{Chen}} \bibnamefont{and}
  \bibinfo{author}{\bibfnamefont{J.}~\bibnamefont{J.Quinn}},
  \bibinfo{journal}{Phys.~Rev.~B.} \textbf{\bibinfo{volume}{\textbf{45}}},
  \bibinfo{pages}{11054} (\bibinfo{year}{1992}).

\bibitem[{\citenamefont{C\^{o}t\'{e} et~al.}(1992)\citenamefont{C\^{o}t\'{e},
  Brey, and MacDonald}}]{cote:92}
\bibinfo{author}{\bibfnamefont{R.}~\bibnamefont{C\^{o}t\'{e}}},
  \bibinfo{author}{\bibfnamefont{L.}~\bibnamefont{Brey}}, \bibnamefont{and}
  \bibinfo{author}{\bibfnamefont{A.~H.} \bibnamefont{MacDonald}},
  \bibinfo{journal}{Phys.~Rev.~B} \textbf{\bibinfo{volume}{\textbf{46}}},
  \bibinfo{pages}{11239} (\bibinfo{year}{1992}).

\bibitem[{\citenamefont{Demler et~al.}(2001)\citenamefont{Demler, Nayak, and
  {Das Sarma}}}]{deml:01}
\bibinfo{author}{\bibfnamefont{E.}~\bibnamefont{Demler}},
  \bibinfo{author}{\bibfnamefont{C.}~\bibnamefont{Nayak}}, \bibnamefont{and}
  \bibinfo{author}{\bibfnamefont{S.}~\bibnamefont{{Das Sarma}}},
  \bibinfo{journal}{Phys.~Rev.~Lett.} \textbf{\bibinfo{volume}{\textbf{86}}},
  \bibinfo{pages}{1853} (\bibinfo{year}{2001}).

\bibitem[{\citenamefont{Veillette et~al.}(2002)\citenamefont{Veillette,
  Balents, and Fisher}}]{veil:02}
\bibinfo{author}{\bibfnamefont{M.~Y.} \bibnamefont{Veillette}},
  \bibinfo{author}{\bibfnamefont{L.}~\bibnamefont{Balents}}, \bibnamefont{and}
  \bibinfo{author}{\bibfnamefont{M.~P.~A.} \bibnamefont{Fisher}},
  \bibinfo{journal}{Phys.~Rev.~B.} \textbf{\bibinfo{volume}{\textbf{66}}},
  \bibinfo{pages}{155401} (\bibinfo{year}{2002}).

\bibitem[{\citenamefont{Yang}(2001)}]{yang:01}
\bibinfo{author}{\bibfnamefont{K.}~\bibnamefont{Yang}},
  \bibinfo{journal}{Phys.~Rev.~Lett.} \textbf{\bibinfo{volume}{\textbf{87}}},
  \bibinfo{pages}{056802} (\bibinfo{year}{2001}).

\bibitem[{\citenamefont{Wen}()}]{xwen:co}
\bibinfo{author}{\bibfnamefont{X.-G.} \bibnamefont{Wen}},
  \bibinfo{howpublished}{private communication}.

\bibitem[{\citenamefont{Cheng}(1981)}]{ycch:81}
\bibinfo{author}{\bibfnamefont{Y.}~\bibnamefont{Cheng}},
  \bibinfo{journal}{Phys.~Rev.~B.} \textbf{\bibinfo{volume}{\textbf{23}}},
  \bibinfo{pages}{157} (\bibinfo{year}{1981}).

\end{thebibliography}

\end{document}